**Scheme of a Derivation of Collapse from Quantum Dynamics II**

by Roland Omnès

Laboratoire de Physique Théorique (Unité Mixte de Recherche, CNRS, UMR 8627), Université de Paris XI, Bâtiment 210, 91405 Orsay, Cedex, France
e-mail: Roland.Omnes@th.u-psud.fr

*Abstract*:

Is wave function collapse a prediction of the Schrödinger equation? This unusual problem is explored in an enlarged framework of interpretation, where quantum dynamics is considered exact and its interpretation is extended to include local entanglement of two systems, including a macroscopic one. This property of local entanglement, which results directly from the Schrödinger equation but is unrelated with observables, is measured by local probabilities, fundamentally distinct from quantum probabilities and evolving nonlinearly. When applied to a macroscopic system and the fluctuations in its environment, local entanglement can also inject a formerly ignored species of incoherence into the quantum state of this system,.

When applied to a quantum measurement, the conjunction of these two effects suggests a self-consistent mechanism of collapse, which would directly derive from the Schrödinger equation.

(This work develops and improves significantly a previously circulated version with the same title [23])

————————

Two momentous papers by Schrödinger [1] and by Einstein, Podolsky and Rosen [2], both published in 1935, left a lasting acceptance that the uniqueness of measurements data would be inconsistent with the Schrödinger equation of evolution. This problem remained since then a matter of worry [3] and is still a subject of much research [4]. It has also become a major theme in the philosophy of science [5].

Quantum physics itself made nevertheless outstanding progress in the meantime. Its laws were always found undeniable foundations for these developments, but the problem of their agreement with a unique macroscopic Reality did not receive a universally agreed answer.

One proposes here a new approach to this problem, according to which the laws of quantum mechanics would be self-consistent and predict wave function collapse as one of their consequences. No revision of the quantum laws themselves would be needed for reaching this result, but the orthodox interpretation, expressed in classical books [6, 7] and used in most textbooks, would be revised significantly. This revision would not be a rejection of the standard interpretation, however, but a broadening making use of a few consequences of the Schrödinger equation, which are still partly conjectural but shed a remarkable new light on the phenomenon of collapse.

## 1. A pattern for a wider interpretation

Many thorough experiments during the last thirty years or so, led to an essential empirical conclusion according to which "wave function collapse" is strictly restricted to macroscopic systems and is never observed in microscopic ones [8]. As a direct consequence and since these systems are never in a pure state, one will dismiss here the usual name of this phenomenon and simply call it "collapse".

Collapse is undoubtedly a physical phenomenon and, moreover, the most frequently observed one since every experiment in quantum physics relies on its systematic and multiple occurrences in experiments. The circumstances under which it happens are well known but one will recall them first here for definiteness:

A quantum measurement is concerned with a microscopic system $A$ and is intended to measure the value of an observable in it. One usually describes it in a case where the initial state of $A$ is expressed by a state vector, which is a superposition of eigenvectors $|k\rangle$ of this observable:

$$.|A\rangle = \sum_k c_k |k\rangle \tag{1.1}$$

This system $A$ interacts with a measuring system $B$, which is always macroscopic. Collapse consists then in the observed fact that a state characterizing a unique value of the measured observable, associated with one of the state vectors $|k\rangle$, comes out from the measurement. Various results occur randomly when the same measurement is performed many times with the same initial state, and the observed frequencies are in perfect accordance with Born's probability law

$$p_k = |c_k|^2. \tag{1.2}$$

A particularly intriguing aspect of collapse is the fact that a measuring device, when it shows off a value of a measured observable, works exactly in the same way as if the initial state of the measured system had involved the unique state vector $|k\rangle$, associated with that value, rather than in the superposition (1.1). One wonders then how a physical effect could be so efficient and universal, and yet be so evanescent that it leaves no sign of its mode of action.

One proposes in this paper that this hidden mode of action relies most probably on local entanglement, which is a property resulting directly from the Schrödinger equation, but also an invisible one because of its lack of relation with observables (*i.e.* self-adjoint operators in Hilbert space [7]).

Section 2 recalls the theory of this effect of local entanglement and extends it somewhat. Section 3 suggests that local entanglement between a macroscopic system and the *fluctuations* in its environment can generate a specific type of incoherence, which would have remained unsuspected hitherto. Section 4 identifies then an explicit effect of "slip in coherence", which acts at the level of a few atoms and would be the elementary agent through which minute transfers occur between the quantum probabilities of various measurement channels. The final Section 5 shows how an accumulation of a huge number of invisible slips of this kind could be responsible for collapse without leaving any trace behind.

The resulting theory leads to drastic revisions and enlargement of the interpretation of quantum mechanics, without any change in its basic dynamical laws. This reconstruction of interpretation is complex and sometimes disturbing by its modifications, so that an attempt at getting clarity will be privileged here rather than a search for rigorous proofs, which would need much harder work.

One may mention that this desire for a minimum of complexity led to a much simplified introduction of local entanglement in Section 2, in spite of the central part of this notion here. A more mathematical approach is sketched in an appendix.

## 2. Local entanglement

Local Entanglement is a direct consequence of the Schrödinger equation. Although one may consider it a genuine consequence of the Schrödinger equation, it shows no relation with quantum observables (*i.e.* no association with self-adjoint operators in Hilbert space [7]). Eliot Lieb and Derek Robinson discovered it in the seventies and gave it this name of "local entanglement" [9]. It drew little attention in the field of interpretation, probably because it was considered mostly as a peculiar effect, specific to many-body physics and not fundamental. The present author rediscovered it in a serendipitous way [10], by ignorance so to say, and called it then "intricacy". One will keep here however its original name of local entanglement, which one will often abridge by "*LE*", even when using this abbreviation to mean "locally entangled" in place of an adjective.

The adjective "local" in this name came in the Lieb-Robinson approach from its association with a spin-lattice model, where the designation of a spin coincides with its location. Although one will recover this association with a location in space, one will rather mean it as associated with individual atoms (or other elementary constituents of a macroscopic system).

The "paradigm" of *LE* with which one will deal consists in a model of a Geiger counter (or a wire chamber, or essentially a gas of atoms in a solid box). It stands then as a well-defined quantum system, which one will denote by *B* and will first suppose isolated. Another system *A*, usually microscopic, can interact with this system *B* and consists in an energetic charged particle, initially in a state (1.1) where the states $|k\rangle$ represent different tracks of the particle.

The same approach holds also when the measuring system involves several separate parts, like in a Stern-Gerlach measurement for instance. These parts can also have eventually a space-like relativistic separation. Although one will discuss mainly local entanglement in the case of a detector made of atoms (for instance a gas of argon atoms acting as a detector and a dielectric in a counter), the discussion will be valid also when excited atoms, ions and free electrons are produced by a charged particle. As a matter of fact, local entanglement depends little on the nature of the particles under consideration and this character contributes to make its discussion easier and general.

One introduces this local entanglement in a simple case where the system *B* is in one piece and the initial state of *A* consists in a unique track, associated with a unique state vector $|k\rangle$. One will also disregard the charge of this particle and represent simply its interaction with atoms by a potential *U*, whereas another potential *V* describes the interaction between pairs of atoms in *B*.

The introduction of *LE* looks then much like a game: One may imagine that, in addition to its quantum behavior, every particle in the *AB* system carries a color, either white or red. Before interaction, the particle $A$ is red and every atom in $B$ is white. One also assumes that the red color is conveyed by contagion so that, when a red particle interacts with a white one, both of them come out red from their interaction. Moreover, when a particle has become red, it keeps that color forever. Finally, when two white atoms interact, they remain white when they come out from interaction.

A mathematical expression of this game consists in replacing the two colors, red and white, by two formal "indices of local entanglement", 1 and 0. The rules of contagion can be expressed then by using three $2 \times 2$ matrices, in which these indices 0 and 1 denote rows and columns, namely:

$$P_0 = \left( \begin{array}{cc} 0 & 0 \\ 0 & 1 \end{array} \right), P_1 = \left( \begin{array}{cc} 1 & 0 \\ 0 & 0 \end{array} \right), S = \left( \begin{array}{cc} 0 & 1 \\ 0 & 0 \end{array} \right). \tag{2.1}$$

$P_0$ can be interpreted as a projection matrix, which picks up an atom with $LE$ index 0 and keeps this index unchanged. The same behavior holds for $P_1$, which picks up Index 1 and conserves it. The matrix $S$ picks up an $LE$ index 0 (which indicates an absence of any previous influence of $A$) and brings it to local entanglement, shown by index 1 (so that the influence of $A$ becomes therefrom imprinted on this atom).

The matrices (2.1) are not meant to act on a state vector in a two-dimensional Hilbert space, but only on a conventional index, which an atom carries. An important feature of this family of matrices is that the matrix $S^\dagger$, which could be formally adjoined to $S$ and would bring back local entanglement to no local entanglement, does not belong to the construction. This absence imposes an irreversible character to local entanglement, in accordance with its representation as a contagion of $LE$.

One can make these rules of contagion mathematically explicit: To do so, one replaces the potential $U_{Aa}$ for the interaction between the particle $A$ and an atom $a$, by a $2 \times 2$ matrix

$$\boldsymbol{U}_{Aa} = U_{Aa} (P_{1a} + S_a) \tag{2.2a}$$

Similarly, the potential $V_{ab}$ for interaction between two atoms $a$ and $b$ is replaced by

$$\boldsymbol{V}_{ab} = V_{ab} (P_{0a} \otimes P_{0b} + P_{1a} \otimes P_{1b} + P_{1a} \otimes S_b + S_a \otimes P_{1b}), \tag{2.2b}$$

which describes adequately the rules of contagion.

This formal construction can be extended easily to the case of several measurement channels $k$, as the ones in (1.1). Every index $k$ is then associated with local entanglement with a definite state vector $|k\rangle$ of $A$ whereas the index 0 represents non-local entanglement (*i.*e. no local entanglement with any channel).

## 2.2. *Dynamics of local entanglement*

One turns then to the dynamical evolution of local entanglement. In the case of a unique channel, for instance $|1\rangle$, the standard wave function $\psi$ of the composite $AB$ system evolves according to the Schrödinger equation

$$i\hbar \partial \psi / \partial t = H\psi . \tag{2.3}$$

When the interaction terms in the Hamiltonian $H$ are replaced by the matrix expressions (2.2), while kinetic terms remain diagonal in $LE$ indices, Equation (2.3) becomes a set of coupled equations for a set of locally entangled wave functions $\{\psi_s\}$: If $N$ denotes the number of atoms in $B$, the index $s$ of a $LE$ function $\psi_s$ is a sequence of $N$ indices of local entanglement 1 or 0.

Initially, all atoms are still non-locally entangled in a unique component with all these indices equal to 0. The Bose-Einstein or Fermi-Dirac symmetry of the wave function $\psi$ remains valid in every $\psi_s$, because two atoms both carry the same index, 0 or 1, when they come out from an interaction.

If one denotes by $\psi'$ this set $\{\psi_s\}$ and one considers it as a vector with $2^N$ components, one gets a linear equation of evolution with the same abstract form as a Schrödinger equation, namely

$$iH'\partial\psi'/\partial t = H'\psi' \qquad (2.4)$$

The operator $H'$ is a $2^N \times 2^N$ matrix. Its matrix elements involve differential operators representing kinetic energy, and potentials ($U$, $V$) representing interactions. Before interaction between the two systems $A$ and $B$, the vector $\psi'$ has only one component in which all the $LE$ indices are 0. This unique component coincides then with the standard wave function $\psi$ and one finds that, because of (2.4), the standard wave function coincides at all times with the sum

$$\psi = \Sigma_s\, \psi_s. \qquad (2.5)$$

Several other properties of $\psi'$ show off on the contrary significant differences in meaning and in form between standard quantum dynamics and local entanglement (although the second one amounts only to rewriting the first one): The evolution operator $H'$ in (2.4) is *not* self-adjoint and, as a consequence (or as the real cause), local entanglement is *irreversible* under time reversal. It always ends up with a situation where all the atomic states have become locally entangled. In the case of several channels, as in the sum (1.1), this final situation coincides with standard entanglement. Moreover, local entanglement stands completely out of the standard interpretation, since no standard observable can extract the $LE$ component $\psi_s(t)$ from the wave function $\psi(t)$ as one of its eigenvectors.

### 2.3. *Probabilities of local entanglement*

One can also construct a quantum field version for local entanglement [10]. This is convenient for extending the domain of $LE$ and draw more of its consequences in macroscopic systems, not only in gases but also in every system that can be analyzed by means of many-body theory and the use of quantum fields [11].

One can construct quantum fields showing local entanglement and denoted by $\phi_r(x)$, where the index $r$ can either be equal to some channel index $k$ in (1.1), or equal to 0 for no local entanglement. The standard quantum field $\phi(x)$ coincides then with the sum of these $LE$ fields  One can also define number densities $n_r(x)$ of locally entangled atoms (or no locally entangled ones) as average values of products $\phi_r^\dagger(x')\phi_r(x')$ over a small space region with center at a point $x$. Although the sum of these averages is not exactly equal to the standard local density $n(x)$ of atoms, it does so with a negligible error when the state of the system is strongly disorganized, as in a gas for instance. This kind of emergence of a classical behavior is well known in statistical physics [11] and one will often use, in accordance of the present work with a first exploration.

One gets thus *local probabilities* of local entanglement, $f_r(x)$, which are defined as the ratios $n_r(x)/n(x)$. They are positive and satisfy the sum property

$$\sum_k p_k f_k(x) + f_0(x) = 1. \tag{2.6}$$

One can interpret this relation as meaning that the atoms near a point $x$ have a probability $p_k$ for being entangled with a channel $k$ in (1.1) and also a probability $f_k(x)$ for being moreover locally entangled with that channel. This set of probabilities is completed by a probability $f_0(x)$ for non-local entanglement.

This existence of local probabilities of local entanglement (and non-*LE*) is the most remarkable outcome of these results, because these local probabilities are not expressible by means of standard observables and do not belong therefore to the standard category of quantum probabilities. One is therefore already trespassing neatly the frontiers of the standard interpretation.

## 2.4. *Propagation of local probabilities of local entanglement*: *Waves of LE*

When Lieb and Robinson discovered local entanglement, they described their properties of propagation as remarkable "light-cone effects", although the corresponding velocity was unrelated with the velocity of light.

The present author considered also these aspects of *LE* in the case of local entanglement between the particle $A$ and a gas of atoms [10]. One will look now at that case, with emphasis on the physical aspects of these effects. One will use again for that purpose a descriptive formulation where there is only one channel and the influence of Particle $A$ can be illustrated as a transmission of color, $A$ being red and communicating this color to initially white atoms, which carry this color farther away.

The collisions between atoms can be considered random and their collective effect is expressed by a probability $f_1(x,t)$ for the atoms near a point $x$ to be locally entangled with $A$ (*i.e.* to be red). Another probability $f_0(x,t)$ is associated with non-local entanglement (or the white color). The two probabilities sum up to 1 almost exactly, namely

$$f_1(x) + f_0(x) = 1. \tag{2.7}$$

An approximation of the evolution of $f_1(x, t)$ by means of classical statistical physics can be justified by the fact that everything in it depends only on random atomic collisions. Similar approximate methods are known significant, at least qualitatively and regarding orders of magnitude, similar transport processes such as heat diffusion or electric conduction [12]. This kind of evolution, which depends only on collisions of atoms, can be described by a diffusion equation

$$\partial f_1/\partial t_{\text{ diffusion}} = D\Delta f_1, \tag{2.8}$$

where $D$ is a diffusion coefficient. It can also be linked simply, as far as order of magnitude are concerned, with the mean free path of atoms $\lambda$ and their mean free time $\tau$ by $D = \lambda^2/6\tau$.

A collision between a non-locally entangled atom and a locally entangled one contributes to the contagion of *LE*. When it occurs near a point $x$ during a short time interval $\delta t$, the associated probability is equal to the product $f_1(x, t) f_0(x, t) \, \delta t / \tau$. The corresponding increase in $f_1(x, t)$ owing to contagion is therefore given by

$$\partial f_1 / \partial t \text{ contagion} = f_1 f_0 / \tau. \tag{2.9}$$

Using (2.8) and (2.9), one gets a nonlinear equation for the evolution and propagation of local entanglement, which is

$$\partial f_1 / \partial t = D \Delta f_1 + f_1 (1 - f_1) / \tau. \tag{2.10}$$

When looking at this equation in a one-dimensional space, one finds that it cannot be satisfied by a function $f_1$, which would be everywhere positive and non-vanishing as it does in the case of the diffusion equation (2.8). In dimension 3, there must exist a moving boundary *S,* which separates a region where $f_1(x, t)$ is positive from a region where it vanishes (this existence of moving fronts is frequent in nonlinear wave equations [13]).

One can get an idea of the motion of the front and of the behavior of $f_1$ by solving numerically this equation (2.10), when it depends only on a one-dimensional variable *x.* The average velocity of atoms is then $\lambda / \tau$ and its average value along one direction of three-dimensional space is $v^{\prime} = 3^{-1/2} v$ (notice that this is the velocity of sound in a dilute gas). Whereas diffusion expands only at time *t* to a distance of order $(Dt)^{1/2}$, diffusion acting together with contagion in Equation (2.10) yields an expansion of *LE* at the much larger distance $v^{\prime}t$, which defines the position of the moving boundary *S* at that time.

Numerical solutions of the propagation equation (2.10) confirm this motion of a wave front *S* at the velocity $v^{\prime}$. The probability of local entanglement $f_1(x, t)$ increases rapidly from zero to 1 behind this front, over a distance of order the mean free path $\lambda$.

### 3. The environment and its interpretation

The second step in the present construction is concerned with the effects of the environment of a macroscopic system. It consists essentially in the following assertion:

**Proposition 1**

*Fluctuations in the action of environment can inject into the state of a macroscopic system a specific form of incoherence, which propagates into the system.*

This effect will be shown a consequence of local entanglement between the macroscopic system and its environment. As long as one uses only the standard interpretation [7], however, one cannot prove the existence of the kind of incoherence in Proposition 1, or express its nature reliably: Two keywords in this proposition, "environment" and "incoherence", do not belong to this interpretation. A third word, "fluctuations", is also external, since it is linked with the notion of "environment", in a sense that does not does not take this environment as a quantum system and is therefore also foreign to the standard interpretation.

A suitable framework for this proposition relies on the "cluster decomposition principle" of quantum theory, which is advocated by Steven Weinberg as necessary for a foundation of quantum field theory on a complete set of principles [14]. This principle can be used also to derive Feynman paths from the principles of quantum field theory ([14], Volume II). A description by Feynman paths can provide a direct approach to the incoherence in Proposition 1: It would be then associated with random phases, originating in external molecules belonging to fluctuations in the environment.

Another aspect of Proposition 1 is concerned with the theoretical status of environment. The question is then whether the environment of a well-defined quantum system can be considered as being also a quantum system.

The previous Geiger counter can be considered well defined, theoretically, in view of its association with a definite Hilbert space and a definite algebra of observables [6, 7]. One might think of "defining" its environment as a wider system surrounding the counter, for instance a definite part of the atmosphere around it, in which case this environment would be described by a grand canonical ensemble. But there would always be a still wider environment around this newly defined environment, with no end except for the whole universe.

One will not adopt this assumption by Everett of the universe as being a perfect quantum system, because of its "many-worlds" unavoidable consequence [15, 16]. One will rather consider that the main consequence of the universe, regarding quantum measurements, is a permanent presence of an environment around any formally well-defined quantum system.

One will consider the environment as an objective datum on which much information is available, but which does not constitute by itself an ideal quantum system.

### 3.1. *The case of a unique molecule and a first axiom of interpretation*

One will use again the example of Geiger counter, still denoted by *B,* in which a solid box encloses a gas of atoms. No measured system *A* is present at the period of time, which one considers now. The environment acting on *B* is supposed to consist only of a limited external atmosphere, which is under standard conditions of temperature and pressure.

To begin with, one considers a unique atmospheric molecule, denoted by *M*, which hits the box and rebounds on it. The previous description of local entanglement (with the outgoing state of *M* in the present case) implies that a wave of local entanglement starts from a point $x_M$ where the collision occurs and expands from there into the counter

One could show more precisely how this collision generates first some phonons, which are locally entangled with the outgoing state of the molecule and begin to propagate local entanglement. This *LE* passes then to other phonons, under a series of phonon-phonon interactions. A description of this propagation by means of Feynman paths (or Feynman graphs), shows that locally entangled phonons can be distinguished from non-locally entangled ones and can be labeled by an index of local entanglement. When the *LE* wave fills the box up, the phase it carries is no more active, because it is present everywhere in all the wave functions of *B*, with no consequence.

The place $x_M$ where *M* hits the box is random, as well as the momentum of the incoming molecule and the momentum transfer $\Delta p$ resulting from the collision. When one considers the initial state of *B* before the collision as an eigenfunction of $\rho_B$, the outgoing wave function of the *MB* system carries a phase $\alpha = x_M . \Delta p / \hbar$, which is also random and is present in all the new eigenfunctions of $\rho_B$ after the collision. This is what one means in Proposition 1 when saying that the environment can inject incoherence into *B*, with the usual meaning of "incoherence" as a presence of random phases.

Another significant datum of this example of a unique molecule, is concerned with the time $\Delta t$ during which a wave of local entanglement crosses the system *B* and keeps its wave functions separated into sums of differently locally entangled ones. This delay is of order $L/c_s$ where $L$ is a typical scale length of the system $B$ and $c_s$ the velocity of the wave (the velocity of sound in this example). One finds this time delay $\Delta t$ of order $10^{-5} L_{cm}$ (in units of one second) if $L_{cm}$ denotes the size of the system $B$ in centimeters.. This is a long time, when compared with the time scales of elementary processes, and this duration will be one the main parameters in the present theory.

### 3.2. *Fluctuations in the action of environment and their theoretical description*

One comes then to the central part of this discussion, which is concerned with the detailed action of environment on the state of $B$. One estimates first some parameters.

Using the rather long time $\Delta t$ during which a wave of local entanglement crosses the system $B$, one can compute how many waves are present in $B$ at an arbitrary time $t$. These waves must have arrived during the time interval $[t - \Delta t, t]$ and their number, which one denotes by $N_t$, is of order $10^{24} L_{cm}^2$ in the present example. In view of the stationary behavior of the system, one can also expect that as many $LE$ waves disappear on average during that time interval, after they filled up $B$ completely.

The fluctuations in these two numbers are of order $N_f = N_t^{1/2}$, or presently $10^{12} L_{cm}$. This is quite large. One knows also that the active part of an $LE$ wave (the region where local entanglement is growing behind its front) has a width of order one mean free path of atom (about $10^{-5} cm$). Various such active regions overlap therefore at every point $x$ in $B$ and their number $N_x$ is of order $10^7$. This is again large and much disorder must be therefore permanently active in a non-perfectly isolated macroscopic system.

One can give a formal expression for this disorder in $\rho_B(t)$, by separating a stable average of this state from its fluctuations: The average action of environment, in the present case, boils down to a pressure acting on the box. It has little interest and one will leave it aside. As far as the gas in the box is concerned, standard methods in statistical physics yield the definite expressions [11]

$$<\rho_B> = Z^{-1} \exp(- H/T), \tag{3.1}$$

where the temperature $T$ is expressed in energy units.

Fluctuations can only belong the difference $\Delta\rho_B(t) = \rho_B(t) - <\rho_B>$. It has a vanishing trace and can be conveniently split into a part $\rho_{B+}(t)$, involving only its positive eigenvalues, and a part $-\rho_{B-}(t)$ involving the negative ones. One gets then

$$\rho_B(t) = <\rho_B> + \rho_{B+}(t) -\rho_{B-}(t). \tag{3.2}$$

The choice of a theoretical description for the environment is a nontrivial problem. As far as its effects on the system $B$ are concerned, one can only get a few data regarding the number of collisions by atmospheric molecules on the external box, during the relevant time interval, as well as the random distribution of their place and time of arrival. Their average effect is only the previously mentioned pressure, and the collisions obey essentially a Poisson distribution.

The part of environment, which can act on $B$ during the interval $[t - \Delta t, t]$, can be restricted to a region of the surrounding atmosphere, which one denotes by $E$ and which is limited by an ideal boundary, at a sufficient distance from the frontier of $B$ for insuring that all the molecules hitting $B$ during that time interval, were always in that region during that time.

From the standpoint of quantum mechanics, one could thus describe the environment as a grand canonical ensemble of molecules, located in this region $E$.

This description holds perfectly well for the action on environment on the system $B$, but not for the reverse effect of $B$ on the state of the atmosphere, which is associated with the return of molecules after collision. A complete quantum account of this coupling between the system $B$ and its environment would require a consideration of a composite system $EB$, with a quantum state $\rho_{EB}$. One excludes this approach because it would lead by extension to a quantum state of the universe.

A proper quantum description of the environment would be a phenomenological representation by a grand canonical ensemble, but although it would allow a quantum description of the action of environment on $B$, it would leave aside the back action of $B$ on its environment. This asymmetric status of the system and of its environment was expressed earlier when one said that an environment is not generally representable by a genuine quantum system.

As a consequence, *the density matrix $\rho_B(t)$ of the system $B$ is a random matrix* [17], by which one means that its matrix elements in a fixed reference system (for instance the eigenvectors of the Hamiltonian of $B$), are random numbers. If so, this behavior is also true, automatically, for the matrices $\rho_{B+}(t)$ and $\rho_{B-}(t)$.

To go farther, one needs a guide and the one we shall use is a guess: Could it be that the randomness of the matrix $\rho_B(t)$, from its environment, could be the one at work when this system $B$ acts in a measurement and undergoes a random collapse?

This is an assumption and, at least at the point where the present theory stands, one will be unable to prove it. This is because a proof needs axioms, and these axioms would have to define an interpretation, which would extend the standard one.

This aim is still too far and one will proceed by means of some remarks and other more guesses, as follow:

One can get an idea of the relevant fluctuations in environment by considering the fluctuations in the number $N_t$ of colliding molecules, which hit the box around B during the time interval $[t - \Delta t, t]$, and also the associated fluctuations in number $N_f$.

One considers a *sample* of these fluctuations, which consist in principle of excesses above the average number of collisions, or deficiencies below, their number being $N_f$. A fundamental property of fluctuations, which are that their samples are intrinsically inaccessible, will be used to pick up at random positive ones, which correspond to excesses, and negative ones corresponding to deficiencies.

Because of the arbitrariness in this construction, one will suppose valuable (in a future theoretical interpretation) a property, which one can establish by looking at a sample, and which is valid for every sample (with anticipation, one may say that this behavior will be found valid for collapse).

Regarding the matrices $\rho_{B+}(t)$ and $\rho_{B-}(t)$, one recalls that in a positive fluctuation by one molecule, the random phase, which is carried by that molecule, passes to an outgoing wave function of $B$ and is absent in what remains of an ingoing wave function. It means that a fluctuating collision either positive or negative, contributes to both $\rho_{B+}$ and $\rho_{B-}$, but these two matrices carry different phases (at least in different places). This behavior will be the main one, which one will need regarding these matrices.

Finally, one notices that the intervening phases (like the previous $\alpha$) have random values, but fixed ones. When one averages on the contrary upon all possible samples, the various phases in various samples behave as a set of absolutely random quantities, in which all of them are independent and every one of them randomly contained in the interval $[0, 2\pi]$

As a last comment, one will consider the "strength" of incoherence, by which one means the value of the common trace $W$ of the two matrices $\rho_{B+}$ and $\rho_{B-}$. One approaches by making assumptions, namely the following ones: (*i*) The action of environment does not spoil appreciably the energy distribution in $\rho_{B}$, as given by (3.1). (*ii*) The external fluctuations are strong enough for making the eigenvectors of a restriction of $\rho_B$ to a small energy interval, randomly oriented with respect to the basis of eigenvectors of $H_B$ (and $<\rho_B>$) in that interval. (*iii*) Some eigenvalues of the perturbed matrix $\rho_B$ can come close to zero.

One can prove that $\Delta\rho_B$ is a Wigner random matrix [17] under these conditions, and the trace $W$ of $\rho_{B+}$ and $\rho_{B-}$ is then equal to its maximal value $4/3\pi$. One will not take this result for granted, but will consider it suggestive enough for assuming that the actual values of $W$ are not extremely small.

*Note*: Some readers could wonder how it could be that such a high amount of incoherence would be present almost everywhere, and was not noticed earlier. The answer is that this incoherence is only present in the matrices $\rho_{B+}(t)$ and $\rho_{B-}(t)$ and their effects cancel in the average value of every observable, which would express an actual *observation*. This incoherence is therefore invisible.

One may mention however that a significant exception exists. It will appear in the forthcoming discussion of collapse, that the probabilities of various measurement channels fluctuate, under the effect of this incoherence. Quite remarkably however, this exception is a confirmation! The reason is that it is concerned with observables belonging to the measured system, and not to the measuring one, in which incoherence holds. One could return the question and say that there could be a unique case where this incoherence would be seen at work, and one sees it everyday in laboratories, where it is called collapse.

### 3.4. *A ballet of LE waves*

When one deals with a definite sample involving a number $N_f$ of fluctuations in external collisions, and one looks at all the associated waves of local entanglement in the matrices $\rho_{B+}$ for instance, these waves look like if they were dancing a ballet. Some of them arose near the beginning of the time interval $[t - \Delta t, t]$, and they had enough time for reaching a wide development in $B$. Other ones occurred near the end of this interval and are still close to the boundary $B$. Most of them are somewhere in-between, with randomly oriented fronts..

Every one of these *LE* wave carries a specific phase, which one denotes again by $\alpha$. This phase is present only behind a moving wave front and absent beyond. It is carried by all atomic states at a distance greater than $\lambda$ behind the front. From this place behind to the front itself, the local probability for an atomic state to carry this random phase decreases gradually from 1 to 0.

In the matrix $\rho_{B+}$ for instance, all these fronts of *LE* waves move around at the velocity of sound. New ones appear permanently on the boundary and other ones disappear, after having filled up the whole system $B$ by their phase. As in Section 2, every one of them is associated with a local probability $f_1(x, t; \alpha)$, which expresses the fraction of atomic states carrying the random phase $\alpha$ near a point $x$. Many fronts of *LE* waves overlap at every point $x$ in $B$ and their number $N_x$, which one already evaluated, is significantly large in the present example.

The number of different random phases, which are carried by different overlapping wave fronts of *LE* waves, is still much larger: In a definite sample of fluctuations, the index *r*, which one used in Section 2 with values 1 or 0 for characterizing local entanglement, is now associated with a definite wave and a definite phase $\alpha$. An eigenfunction of $\rho_{B+}$, in the case of a definite sample of fluctuations, is associated with a number of $LE$ waves equal to $N_f$ and an equal number of associated phases. Near a point $x$, there are about $N_x$ different waves, and so many indices of local entanglement, which are either equal to 1 or 0 (this number is systematically 1 far enough behind the front, and systematically 0 afore).

When one goes from one sample to the set of all samples, the phases occurring near a point $x$ keep their number $N_x$, but become undetermined in the interval $[0, 2\pi]$.

One can express this situation by two significant propositions, which are as follow:

**Proposition 2**

Every eigenvector (or eigenfunction) $\psi$ of a matrix $\rho_{B+}(t)$ splits under the effect of fluctuations in environment into a sum of component wave functions

$$\psi = \sum_n \psi_n, \tag{3.3}$$

where every component $\psi_n$ is a wave function carrying a specific phase and does not extend in space over a distance greater than a mean free path of atoms. Every component $\psi_n$ involves at most a limited number of atoms, of order

$$N_c = n_a \, \lambda^3 . \tag{3.4}$$

**Proposition 3**

The contributions to the matrix $\rho_{B+}(t)$ of two space regions inside the macroscopic system $B$, which are separated by a distance larger than an atomic mean free path, are independent.

The same propositions hold of course for the matrix $\rho_{B-}(t)$. Proposition 3 can be expressed by considering explicitly two space regions $R$ and $R'$ in $B$. They are associated with two local density matrices, $\rho_{R+}$ and $\rho_{R'+}$, which are defined respectively by partial traces of $\rho_{B+}(t)$ over the atoms outside of $R$, or outside of $R'$. The proposition results from the fact that these two matrices involve unrelated components, which carry different random phases. One can express this property mathematically by introducing the union of the two regions $R$ and $R'$. One has then

$$\rho_{R \cup R'} = \rho_R \otimes \rho_{R'} . \tag{3.5}$$

**4. Slips in coherence**

A derivation of collapse begins then by pointing out an elementary mechanism, which will be considered responsible for generating the phenomenon of collapse. One will call this element a "slip in coherence": It consists in a very small alteration in the conservation of quantum probabilities, when two atoms collide under specific conditions.

One must take into account that this phenomenon occurs when the system $B$ is interacting with the microscopic system $A$ during a measurement. The state of $B$ is still under the permanent influence of fluctuations in its environment, and involves a high amount of incoherence. The initial state of the measured system $A$ is supposed given by the superposition (1.1). Local entanglement between the two systems $A$ and $B$ begins as soon as they interact.

A slip in coherence consists then by definition in a collision between two atoms, say $a$ and $b$, under the following conditions:

($i$) The collision is incoherent.

($ii$) The state of Atom $a$ is locally entangled with a state $\left| j \right\rangle$ of the system $A$.

($iii$) The state of Atom $b$ is non-locally entangled with the system $A$.

The ($a$, $b$) collision is governed by the composite density matrix $\rho_{AB}$, which can be decomposed as in (3.2) into the sum of an average and of two components, $\rho_{ABi+}$ and -$\rho_{AB-}$, with opposite signs. Condition ($i$) restricts the slip to a collision that is governed by these last two matrices. One restricts first attention to $\rho_{AB+}$.  In view of entanglement between the systems $A$ and $B$, Equation (3.2), which expresses one of its eigenvectors, becomes

$$\psi_{AB} = \sum\nolimits_{n,k} \left| \psi_{Bnk} \right\rangle \otimes \left| k \right\rangle. \tag{4.1}$$

It will be convenient, for avoiding long discussions, to consider that the various components $\psi_{Bnk}$ have the same random phase for various indices $k$ of entanglement and the same index $n$ denoting a specific set of phases in a sample of collisions. The absolute values of the various coefficients $c_k$ in (1.1) were absorbed for convenience in Equation (3.6) into the norms of the associated components $\psi_{Bnk}$.

In view of Condition ($ii$) and the fact that local entanglement with a state of $A$ implies algebraic entanglement with that state, the state of Atom $a$ belongs necessarily to some function $\psi_{Bnj}$. The $ab$ collision can happen sometimes to be coherent, but only when the state of $b$ belongs to a wave function $\psi_{Bnk}$ showing the same index $n$ characterizing the same random phase. The same quantum state of $b$ is present then in every component $\psi_{Bnk}$ for every index $k$, because of Condition ($iii$), which requires its non-local entanglement.

Conversely, when the state of Atom $b$ carries a phase index $n' \neq n$, the collision is incoherent. In view of the very large number of these indices $n'$, one can assert that the number of coherent $ab$ collisions is negligible with respect to the number of incoherent ones. Condition ($i$) is therefore valid for most collisions and slips in coherence are very frequent events.

One may consider now these slip events: The state of Atom $b$ carries then a phase, which is random with respect to the phase of the state of Atom $a$. All the matrix elements of an $ab$ collision vanish then under averaging on this relative random phase. Since algebraic entanglement is a linear property, which requires a unique global phase in the wave function where it occurs, it loses its power of selection when there is incoherence. All the matrix elements of a collision matrix vanish then when one sums over all possible samples of collision, because it makes phases absolutely random and not only with different values $n$ and $n'$ in different components $\psi_{ABn}$ like the ones in (3.3)).

The conclusion is opposite regarding the squares of matrix elements for a collision, because they do not carry the phases of incoming states: They are insensitive to averaging on random phases. Moreover, these squares are identical for all indices $k$ of algebraic entanglement of $b$ with the states various states of $A$ with indices $k$, because of the absence of local entanglement of Atom $b$. The slip becomes then a full-fledged contagion of the complete state of Atom $b$ to local entanglement with the state $\left| j \right\rangle$ of $A$, and accordingly a switch of the full outgoing state of the collision towards algebraic entanglement with this state $j$.

An essential consequence of this slip is a generation of small variations $\delta p_k$ in the quantum probabilities of the various channels. The calculation yields explicitly

$$\delta p_j = + \, W \, p_j \, (1 - p_j) \, f_j \, (x) \, f_0 \, (x) / 2N_c, \tag{4.2a}$$

$$\delta p_{j'} = - \, W \, p_j \, p_{j'} \, f_j \, (x) \, f_0 \, (x) / 2N_c, \qquad \text{for } j \neq j'. \tag{4.2b}$$

The factor $W$, as well as the signs in these equations, express that the collision is governed by the matrix $\rho_{AB+}$ (all the signs are opposite in the case of $-\rho_{AB-}$). The factor $f_j(x)$ is the probability for validity of Condition (*ii*) and $f_0(x)$ does the same for Condition (*iii*). One recalls that the notation $x$ denotes the place in $B$ where the collision occurs.

One can extend the domain of validity of these results to more realistic phenomena in actual measurements: One considered here only the collisions between atoms, and there are such events in a gas acting as dielectric in a Geiger counter. Free electrons and ions are produced by a charged particle, free electrons are accelerated by an electric field, and so on. But there is no essential difference regarding local entanglement: It works in the same way with neutral atoms, excited ones, ions and electrons, even eventually with photons (from the decay of excited atoms). One will mention later on relevant orders of magnitude but, presently, regarding only matters of principles and of consistency, one may say that slips in coherence could be essential agents in collapse, since they provide simply an answer to one of the main associated questions: "How can there be variations in the quantum probabilities of various measurement channels?". (The author looked of course at a variety of other measurements, if only to check whether some of them would produce obvious counterexamples. This review raised interesting new problems, but no obvious counter-evidence. One will leave it aside here).

A difficulty could have been linked with the non-separable character of quantum mechanics, particularly when a measurement uses several separate detectors, like in a Stern-Gerlach experiment. The necessary change in the present approach is obvious and purely formal. It amounts simply to extend the domain of definition of the position variable $x$ in Equations (4.2) to the union of all space regions inside the detectors. An adaptation to other different parameters in various detectors, or various places in one of them, is trivial/

## 4. Collapse as a quantum phenomenon

The theory of collapse becomes almost straightforward when one uses Equations (4.2). Its mechanism relies on an accumulation of transitions in quantum probabilities, which result from all the slip events entering among all the atomic collisions during a short time $\delta t$. One must of course consider also the effects of the two matrices $\rho_{AB+}$ and $-\rho_{AB-}$. Everything boils down to sum the results of equations such as (4.2), with variants taking account of various states $j$ entering in them and of all the places $x$ where collisions occur between unexcited atoms or other particles.

One will only consider two atoms (or call "atoms" particles participating in a slip). One will not enter in detailed calculations, which are straightforward, and only look at a few aspects of their results.

When doing these calculations, one compares first Equations (4.2) with the same ones holding under slightly different conditions. One dealt for instance with Atom $a$ in the case where it was locally entangled with Channel $j$ and gave rise to small transfers of quantum probabilities from the channels with index $j' \neq j$, towards this channel $j$. There are other slips, where the atom playing the part of *a* is locally entangled with one of these channels *j'*: The average variations in probabilities, $\delta p_i$ and $\delta p_{j'}$ cancel each other in these two cases (in view of the symmetry of the right-hand side of (4.2b) in the indices *I* and *j*. The standard deviations $<(\delta p_j)^2>$ as well as the correlation coefficients $<\delta p_j \delta p_j>$ do not vanish however and they even add up. When one considers the matrix $-\rho_{AB-}$ after having dealt with $\rho_{AB+}$, the results again add up.

One gets thus the final results

$$\left\langle \left(\delta p_j\right)^2 \right\rangle = W p_j \left(1 - p_j\right)(\delta t / \tau)\frac{1}{N_c^2}\int n_a f_j(x) f_0(x) dx, \tag{5.1}$$

$$\left\langle \delta p_j \delta p_k \right\rangle = -W p_j p_k (\delta t / \tau)\frac{1}{N_c^2}\int n_a [f_j(x) + f_k(x)] f_0(x) dx, \quad \text{for } j' \neq j. \tag{5.2}$$

(The local probability $f_0(x)$ for no local entanglement is again given in these expressions by Equation (2.6))

## 5.1. *From fluctuations to collapse*

The linear behavior in $\delta t$ of the correlations (5.1-2) implies that the set of random quantum probabilities $\{p_i\}$ undergoes a Brownian random process. Philip Pearle suggested rather long ago the possible essential relevance of these processes in collapse [19]. Because of Schrödinger's no-go conclusion [1], which was undisputed (till now), he considered logically that an occurrence of this kind of process would require violation of the Schrödinger equation. A more recent theory of "continuous spontaneous localization" (CSL) has extended more recently Pearle's results [19, 4], by a combination with the Ghitardi-Rimini-Weber assumption of a physical effect, which would adding a random action to the evolution under Schrödinger's equation [20]. One does not need here this GRW effect and one considers Schrödinger's equation as "the" unique Law of quantum dynamics.

The essential of Pearle's conception stands on a key theorem, which he proved in various ways: According to this theorem, a Brownian random process leads unavoidably to a collapse effect: The various quantum probabilities of most channels vanish successively, until a unique one (say for instance $p_j$), reaches the fatidic and final value 1. It turns out (and this is the beauty of this theorem) that the Brownian probability for this outcome is identical with the initial value $\left|c_j\right|^2$ of this quantity $p_j$, in perfect agreement with Born's fundamental law.

Presently, one must look carefully at the conditions of validity for Pearle's theorem. They consider the fluctuations as random, infinitely small and infinite in number. When one introduces accordingly a probability distribution $\Phi(p_1, p_2,...; t)$ for the random quantities $\{p_j\}$, it must satisfy the Fokker-Planck equation

$$\partial\Phi/\partial t = \sum_{jj'}\partial_j\partial_{j'}\{<\delta p_j \delta p_{j'}> \Phi\}, \tag{5.3}$$

with initial conditions

$$\Phi(0, \{ p_j \}) = \prod_j \delta(p_j - | c_j |^2). \tag{5.4}$$

The relevant boundary conditions are that $\Phi$ vanishes on the boundary of the $\{p_j\}$ domain (defined by $p_j \geq 0$, $\Sigma_j\, p_j = 1$): This boundary consists in various parts, defined by the equations $p_j = 0$ for every value of the index $j$.

The assumptions of the theorem require that a quantity $p_j$ can actually reach the value 0, so that the associated channel can disappear. This condition can be expressed explicitly by introducing a Fokker-Planck probability current $J$, with components

$$J_j = \partial_{j'}\{<\delta p_j\ \delta p_{j'}> \Phi\}\ . \tag{5.5}$$

Pearle's theorem requires that the component $J_j$ of this current does not vanish on the parts of the boundary where some $p_j$ is zero. This is necessary for allowing $p_j$ to vanish and getting a finite value for the average time of collapse (otherwise, collapse would take an infinite time…). It seems at firs sight that there is a difficulty there, with the correlation coefficients in (5.1-2): They give

$$J_j = \{\partial_{j'}<\delta p_j\ \delta p_{j'}>\}\Phi + <\delta p_j\ \delta p_{j'}>\partial_{j'}\Phi, \tag{5.6}$$

where $\partial_{j'}$ is meant as $\partial/\partial p_{j'}$. The first term vanishes on the boundary because of the boundary condition $\Phi = 0$ for $p_j = 0$. The second term vanishes also in view of the explicit dependence (5.1-2) on the $p_k$'s, including the expression of the quantity $f_0(x)$).

One is thus led apparently to a sad conclusion, which would be that no randomly varying quantity $p_j$ would ever be able to vanish: Schrödinger's analysis would eventually need some revision, its essential conclusion regarding the impossibility of collapse would remain.

It may be worth mentioning that this impediment came only to attention at the last step in the present research, like if one had been hunting for the snark and got a boojum [21]. The relieving answer came only after a few days, much like a "deus ex machina" last event in a play. This is how it goes:

The Fokker-Planck equation relies on infinitely small random variations of a purely mathematical nature. But individual variations are finite in the present theory: they are due to a rather large number of slips during a short time $\delta t$ (for instance the duration of a two-atoms collision). Every slip yields the finite effects (4.2). If one covers the space in the system $B$ into a lattice of cells with size $\lambda$, the Bose-Einstein's or Fermi-Dirac's indistinguishable character of atoms implies that every individual slip is entirely characterized by the cell $\beta$ where it occurs and the channel $k$ with which there is initially local entanglement (also whether the collision is positive or negative, occurs in $\rho_{AB+}$ or $\rho_{AB-}$).

In view of Proposition 3, the finite variations in $p_j$ from the slips in different cells, add up, so that these local effects can have a significant global action. There is also another much less obvious magnifying effect: If one denotes by $N_{k\beta}$ the number of slips of a given type during a given time interval, this is a random integer and it has a very small average value $<N_{k\beta}>$. The magnifying effect comes then from the Poisson distribution of the values of t $N_{k\beta}$, together with the famous property, which makes this distribution sometimes call the "law of small numbers", namely: The expression $\Delta N_{k\beta} = (<N_{k\beta}>)^{1/2}$ for fluctuations in this distribution implies that the standard deviation $\Delta N_{k\beta}$ is much larger than the average $<N_{k\beta}>$, when this average is much smaller than 1. These conditions are satisfied here. There is accordingly a strong magnification of the fluctuations $\delta p_j$ when $p_j$ is small. It means that a quantum probability can vanish under the effect of a finite number of conspiring slips, or even a unique one. If the Fokker-Planck equation had been exact, as in CSL theories, the time scale of collapse would have been predicted infinite by Equations (5.1-2). Rough estimates confirm this standpoint, but one will leave its details for future more quantitative studies.

**5.2** *Ultimate collapse*

The main remaining question, which readers could wait for, is concerned with quantitative estimates. Only a rough one will be proposed: In the model of a Geiger counter with which one dealt: Formula (5.1) yields a time scale $\tau_c$ of collapse, of order

$$\tau_c \cong \tau \left( n_a \lambda^5 / L^2 W \right), \tag{5.7}$$

where $n_a$ denotes the number density of atoms in the gas. A time scale of collapse of the order of $10^{-10} s$ comes out from this estimate as indicative.

More careful considerations regarding an actual detector could imply a significant increase in efficiency: If one denotes by $\Delta$ the size of the cloud of free electrons in an actual detector when ionization is progressing, ,one may expect a decrease of the time scale (5.7) by a factor of order $\lambda/\Delta$. The estimate looks then sensible.

A further look at matters of principle draws out a more surprising possibility, which could go as far as making the concept of time scale empty. It is inspired by Wojciech Zurek's attractive proposal of "quantum Darwinism" [22] and is concerned in the present case with an eventual presence of other organized systems, which would stand outside the measuring device and would in some sense "observe" it (like an electric current or a microprocessor can be said to "observe" a detector). These systems would participate in the collapse process and would strongly enhance it, making it much shorter. The outcome would be surprising, from a philosophical standpoint regarding Reality [5]: When a unique measuring channel would come out at last, all the past histories of random evolution in other channels, which would have occurred in the meantime, would be wiped out forever into definitive oblivion. What happened during collapse, in the measuring device and its helpers would leave absolutely no trace and no memory.

Except for science fiction writers, the real result with scientific and philosophical value of this study is that collapse would be not only a special and very important consequence of quantum dynamics, but also that its lack of connection with observables means that, intrinsically, as a consequence of the quantum principles, its working would be fundamentally inaccessible by experimental methods.

One will not draw more conclusions, except for saying again that this theory of collapse is proposed only as a conjecture, and remains in wait for really thorough investigations. To which one will add however that the present ideas —whatever their value—seemed able to raise new possibilities and shed unexpected light on old problems. Although that does not imply in any way that the proposed conjecture is true, it makes one believe that something deep and true could exist along that direction. The next step would not be then so much to give proofs, which one can presume hard, but to construct first a really new interpretation of quantum mechanics.

### Acknowledgements

I am particularly thankful to Philippe Blanchard, Robert Dautray, Jürg Frölich, Phillip Pearle and Jean Petitot for their encouragements during the long period of this research. Inspiring exchanges with Franck Laloë were most helpful for clarifying these ideas. I owe also much to the memory of Bernard d'Espagnat who followed this project encouragingly, with close friendship, almost until its completion.

### Appendix: Formalism of Local Entanglement

One indicates in this appendix a few mathematical aspects of local entanglement, in the case of a unique measurement channel: In the standard interpretation of quantum field theory, one uses creation and annihilation field operators, $\phi^\dagger(x)$ and $\phi(x)$ for atoms obeying Bose-Einstein statistics. A quantum vector state of the system $B$ is written as

$$| \psi \rangle = \int dx_1 \, dx_2 \, \dots \, dx_N \, \psi(x_1, x_2; \dots x_N) \phi^\dagger(x_1) \dots \, \phi^\dagger(x_N) \, | \, 0 \, \rangle. \tag{A.1}$$

The fields $\phi$ and $\phi^\dagger$ obey the commutation relations

$$[\phi(x), \phi^\dagger(x')] = \delta(x - x'). \tag{A.2}$$

Locally entangled quantum fields $\phi_r(x)$, with index $r = 0$ or 1, are defined as

$$\phi_r(x) = P_r \, \phi(x), \tag{A.3}$$

where $P_r$ is one of the matrices in (2.1). Their commutation relations are

$$[\phi_r(x), \phi_{r'}(x')] \; = 0,$$

$$[\phi_r(x), \phi_{r'}{}^\dagger(x')] = \delta_{rr'} P_r \, \delta(x - x'), \tag{A.4}$$

Another pair of field ($\alpha(x)$, $\alpha^\dagger(x)$) is used also for describing the particle $A$. with which the gas is entangled. An evolution operator $H'$ propagates $LE$ and is given by

$$H' = H_{A0} + H_{B0} + V_{AB} + V_B, \tag{A.5}$$

where

$$H_{A0} = \int dy \, \alpha^\dagger(y)(- \nabla^2/(2m_A))\alpha(y), \tag{A.6a}$$

$$H_{B0} = \int dx \, \{\phi_1{}^\dagger(x)(- \nabla^2/(2m_a)) \, \phi_1(x) + \phi_0{}^\dagger(x)(- \nabla^2/(2m_a)) \, \phi_0(x)\}, \tag{A.6b}$$

$$U_{AB} = \int dx dy\; \alpha^\dagger(y)\; \phi_1^{\;\dagger}(x) U(x,y)(\phi_1(x) + \phi_0(x))\; \alpha(y), \tag{A.6c}$$

$$V_B = (1/2) \int dx dy\, \{\phi_0^{\;\dagger}(x)\; \phi_0^{\;\dagger}(x')\; \phi_0^{\;\dagger}(x) V(x,x')\; \phi_0(x)\; \phi_0(x')$$

$$+ \phi_1^{\;\dagger}(x)\; \phi_1^{\;\dagger}(x')\; \phi_0^{\;\dagger}(x) V(x,x')\; \phi_1(x)\; \phi_1(x')$$

$$+ \phi_1^{\;\dagger}(x)\; \phi_1^{\;\dagger}(x')\; V(x,x')\; \phi_1(x)\; \phi_0(x')$$

$$+ \phi_1^{\;\dagger}(x)\; \phi_1^{\;\dagger}(x')\; \phi_0^{\;\dagger}(x) V(x,x')\; \phi_0(x)\; \phi_1(x')\}. \tag{A.6d}$$

The fields $(\phi_r(x),\; \phi_r^{\;\dagger}(x))$ are *not* operators in the Hilbert space where $\phi(x)$ is an operator (so that real quantities such as $\phi_r(x) + \phi_r^{\;\dagger}(x)$ or $\phi_r(x)\phi_r^{\;\dagger}(x)$ are not observables in a usual sense). A more complete mathematical formulation would make these fields act as operators inside a *sheaf* of Hilbert spaces, in the sense of sheaf theory where a field $\phi_r(x)$ for instance, would bring transitions from one *LE* Hilbert spaces to other ones, like an analytic function allows transitions from one complex plane to other ones in a Riemann surface. A wave function $\psi_s$, where *s* is a sequence of *LE* indices of atoms, would belong to a specific Hilbert space in a sheaf. This formulation in terms of sheaves has not been attempted however and the present approach relies still more on intuition (or on formalism) that on rigorous mathematics.

*Dynamics of local entanglement*:

Before interaction of the two systems *A* and *B*. a state vector of *B* is expressed as in Equation (A.1), with non-locally entangled fields $\phi_0^{\;\dagger}(x)$ replacing the standard fields $\phi^\dagger(x)$. The operator *H'* in (A.5) can act on this initial state as soon as the *AB*-interaction begins.

The terms in Equations (A.6c) and (A.6d) describing switches in local entanglement yield cluster decompositions of the standard wave function $\psi_{AB}(t)$ into a sum of locally entangled ones, $\psi_s(t)$, as soon as the two systems interact. Similarly, because a macroscopic system of actual interest is never in a pure state, one must describe it by a density matrix $\rho_{AB}(t)$, which becomes a locally entangled matrix $\sigma_{AB}(t)$ under the construction of locally entangled wave functions.

The equation describing the evolution of the locally entangled matrix $\sigma_{AB}$ is then

$$i\hbar\partial\sigma_{AB}\,/\,\partial t = H'\sigma_{AB} - \sigma_{AB}H'^{\dagger} \tag{A.7}$$

*Local probabilities of entanglement*

The local probabilities of *LE* and non-*LE*, $f_1(x)$ and $f_0(x)$, are constructed as follow: One introduces first two operators, $N_1$ and $N_0$, for the total numbers of locally entangled and non-locally entangled ones, namely

$$N_r = \int \phi^\dagger_{\;r}(x')\phi_r(x')dx', \tag{A.8}$$

where the integral extends on the whole space of *B* and the *LE* index *r* can take the values 1 or 0. One can also introduce operators $N_{rC}$ for the number of locally entangled atoms or non-locally entangled atoms in a space cell *C*, by using (A.8) with the integral on *x'* restricted to that cell. If one denotes by $x$ the center of the cell, one can define formally two local probabilities of *LE* and non-*LE*, $f_1(x)$ and $f_0(x)$, by

$$f_r(x) = Tr(N_{rC}\,\sigma_{AB})/\,N_C\ ,\tag{A.9}$$

where $N_C$ denotes the average number of atoms in the cell $C$.

These quantities are positive and can be used as local measures of entanglement, as done in Section 2. The sum property (2.7), which makes them meaningful as probabilities, becomes then

$$Tr[\,\sigma_{AB}\,\{\int \phi^\dagger_1(x)\phi_1(x)dx + \int \phi^\dagger_0(x)\phi_0\,(x)dx\,\}] = Tr[\,\sigma_{AB}\int \phi^\dagger(x)\phi(x)dx]\tag{A.10}$$

with all the integrals extending on the cell $C$. The condition of validity for (2.7) is then

$$Tr[\,\sigma_{AB}\int (\phi^\dagger_1(x)\phi_0(x) + \phi^\dagger_0(x)\phi_1(x))\,dx] \approx 0.\tag{A.11}$$

One assumes that this property holds because of several conjugated reasons: The number of eigenfunctions of the restriction of the matrix $\rho_{AB}$ or of $\sigma_{AB}$ to a cell $C$ is extremely large (of exponential order in $N_C$). The trace (A.9), which involves this restriction of $\sigma_{AB}$, is a sum, over all pairs of these eigenfunctions, of so many real terms. As in many-body theory [11], one considers the contribution to (A.11) of the sums of all pairs of different eigenfunctions as negligible, because they involve an extremely high number of real quantities with positive and negative signs. One may considered this sum as negligible, when compared to the sum on diagonal terms, which enter in (A.11).

Finally, one acknowledges the very sketchy character of these indications regarding a mathematical formalism: they have at least the interest of showing how much would have to be done for insuring a valid explanation of collapse, if it made sense along the proposed direction.